# Long-Term Research & Design Strategies for Fusion Energy Materials

David Cohen-Tanugi[1], Myles G. Stapelberg[2,3], Michael P. Short[2,3], Sara E. Ferry[2], Dennis G. Whyte[2,3], Zachary S. Hartwig[2,3], and Tonio Buonassisi[4]

[1]Proto Ventures, Massachusetts Institute of Technology, Cambridge, MA 02139, USA

[2]Plasma Science & Fusion Center, Massachusetts Institute of Technology, Cambridge, MA 02139, USA

[3]Department of Nuclear Science & Engineering, Massachusetts Institute of Technology, Cambridge, MA 02139, USA

[4]Department of Mechanical Engineering, Massachusetts Institute of Technology, Cambridge, MA 02139, USA

## Progress and potential

- Fusion energy is entering a new stage of development, with recent successful scientific demonstrations and multiple pilot plants planned in the next decade.

- Beyond the pilot stage, the materials required for commercially viable fusion power plants do not exist today. This represents a unique opportunity for materials scientists to quickly become involved in industrially relevant fusion materials R&D that spans from near-term pilot devices to long-term power plant fleets.

- A harsh physical environment gives rise to materials challenges across many components including plasma-facing components, magnets, blankets, but also less-studied subsystems such as those for the fuel cycle, heat exchangers, sensors, insulating components, functional materials, and maintenance.

- The challenge of fusion materials is compounded by the fact that design targets are not precisely defined and are rapidly evolving, and by a lack of representative testing facilities. Iterative co-evolution design can help materials scientists maximize their impact in this high uncertainty technology environment.

- The innovation process can be significantly accelerated by meticulously studying the evolution of fission materials science, anticipating challenges such as outage reduction or the coupled effects of radiation, corrosion, and mechanical stress. This proactive approach will ensure that the first fusion power plants will experience fewer materials degradation issues than the first generation of fission plants did decades earlier.

# Summary

Fusion energy is at an important inflection point in its development: multiple government agencies and private companies are now planning fusion pilot plants to deliver electricity to the grid in the next decade. However, realizing fusion as a technically and economically viable energy source depends on developing and qualifying materials that can withstand the extreme environment inside a fusion power plant. This Perspective seeks to engage the broader materials science community in this long-term effort. We first outline the principal materials challenges and research opportunities for fusion. Next, we argue that fusion is distinct from other energy applications with respect to materials, not just in the magnitude and complexity of the technical challenges but also in the present level of uncertainty in materials design requirements. To address this, we finally propose a research framework based on an iterative co-evolution of materials science and fusion power plant design requirements.

***Keywords***: *fusion, energy, materials science, tritium, design uncertainty, co-evolution, self-driving labs*

1. **Introduction**

Fusion holds the promise of firm energy production without operational $CO_2$ emissions and has recently reached an important inflection point in its development towards a commercial energy source[1]. Most visibly, the US National Ignition Facility has twice recently demonstrated net energy gain in a research setting[2]. The past few years have also seen the emergence of a budding commercial fusion industry: working alongside several publicly funded programs worldwide[3,4,5], over a dozen private companies are now investing billions of dollars to develop fusion pilot plants designed to deliver power to the grid in the early 2030s, with commercial plants planned later in that decade[6,7]. Thus, the field of fusion energy is now shifting into high gear, with faster timelines motivated both by societal-environmental objectives as well as private sector investments.

However, significant materials challenges stand between the current state of the field and what is required for commercial fusion power plants. While early pilot plants can be regarded as feasible with existing materials, it is generally acknowledged that the materials and components necessary for economically viable fusion power plants — which must operate over several decades without significant downtime — have not yet been developed. This is because the working environment for materials inside a fusion power plant is uniquely harsh, resulting in materials challenges that far exceed those for other energy technologies, including nuclear fission reactors.

These developments represent an unprecedented opportunity for materials scientists looking to become involved in fusion energy. The magnitude of the technical challenge and the fast pace of commercial development, paired with the new availability of several important sources of R&D funding, means that the field of fusion materials, formerly a niche branch of nuclear materials, now presents many research opportunities to a broader cross-section of the materials science community.

In this Perspective, we outline key materials challenges facing the emerging fusion power sector, and we show how the wider materials community can more actively participate in this research. Importantly, we seek to distinguish fusion from other clean energy materials challenges not just in the degree of technical complexity, but also in the high level of uncertainty in fusion plant concepts and in design requirements. We discuss approaches to materials R&D that are well suited to this high-uncertainty environment, and place emphasis on design co-evolution and convergent thinking. We argue for this new and faster way of improving existing materials, discovering novel materials, developing advanced manufacturing techniques, growing material supply chains, and establishing new facilities focused specifically on coupled effects for high-fidelity assessment of leading material candidates.

Given the concentration of commercial organizations and research institutions pursuing fusion in the United States and the geographical location of the authors, this article primarily represents a US-centric perspective on materials needs and opportunities for fusion energy. For simplicity, we limit our discussion to magnetic confinement fusion (MCF) devices using the deuterium-tritium (D-T) fusion reaction. These devices presently provide the most technically mature environments for understanding fusion material requirements and challenges. Of course, many elements of our approach should apply qualitatively to other fusion concepts as well, such as magneto-inertial and inertial fusion, which have partially-overlapping yet distinct set of materials challenges from MCF.

## 2. Selected materials challenges in commercially advanced fusion

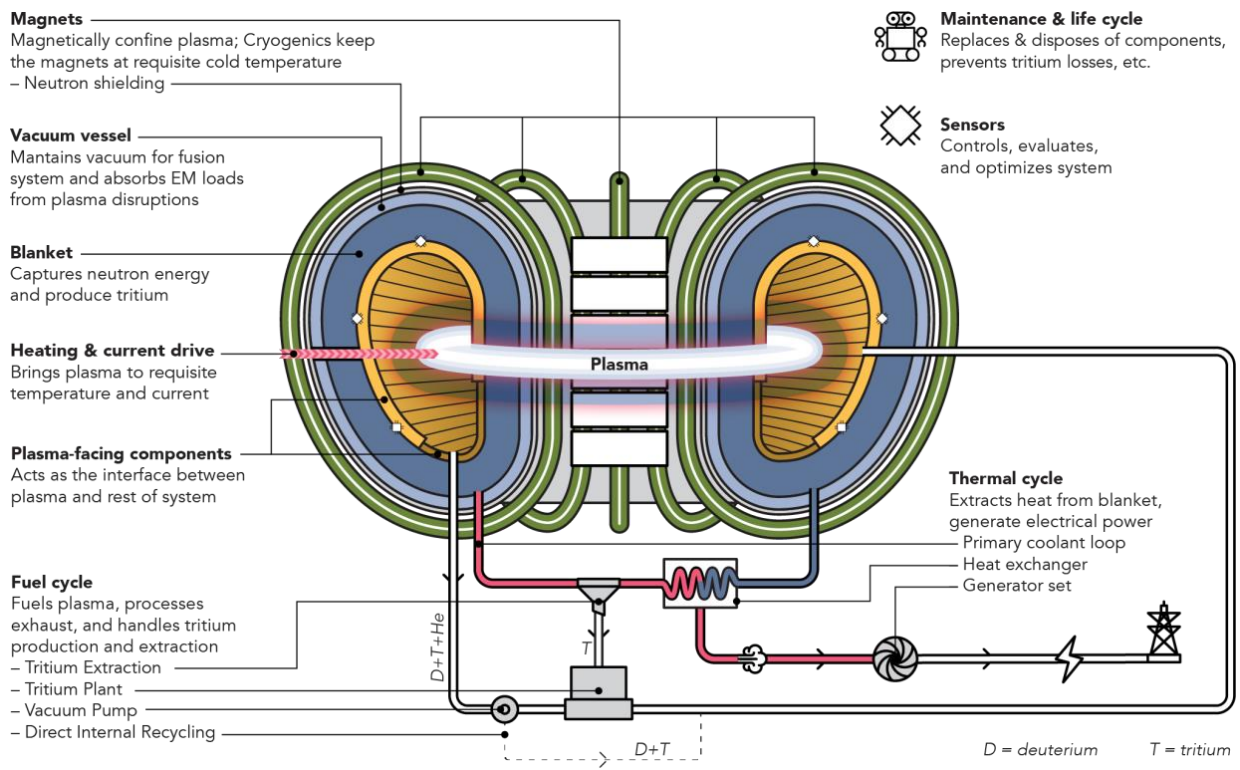

*Figure 1: Functional anatomy of a fusion plant with deuterium-tritium fuel and magnetic confinement. A wide variety of fusion plant designs have been proposed. A generic but representative tokamak design is shown here for illustrative purposes. The key functional systems are present in other tokamak designs and in most other non-tokamak concepts as well.*

At the highest level, a magnetically confined D-T fusion electric power plant produces electricity from the fusion reaction of deuterium and tritium in a high-temperature plasma. This requires, *inter alia*, magnets to confine the plasma, a blanket to convert radiation into heat and to generate tritium, a thermal cycle to convert heat into electrical power, and a fuel cycle to continually fuel the plasma with more deuterium and tritium and to manage plasma exhaust (see Figure 1). In some designs, a single component serves multiple functions, such as liquid immersion blankets that act both to produce tritium and cool the power plant. For greater detail, several in-depth references review the fundamentals of fusion energy systems[8][9]. Several landmark consensus reports by the fusion community describe critical research gaps facing the fusion industry[10][11][12][13]; in particular, a subset of materials challenges close to the plasma are well understood and have been described in established reviews[14][15][16][17], while a broader set of materials challenges are only now coming to light in the research community, especially further from the reactor wall.

**Why is fusion such a hard materials problem?** A fusion power plant faces many of the same materials challenges as other industrial-scale energy sources, including manufacturing and mechanical loading of complex geometries, corrosion and cracking, manufacturing tolerances and assembly, and so on; however, it is the unique operating environment of a fusion power plant that separates fusion from other applications of materials. In general, challenges for fusion-materials development arise from both (i) a uniquely harsh operating environment with combined stressors, and (ii) a lack of representative laboratory-scale environments to test candidate materials and validate computational predictions in the multi-effect environments that are expected to be present in fusion power plants.

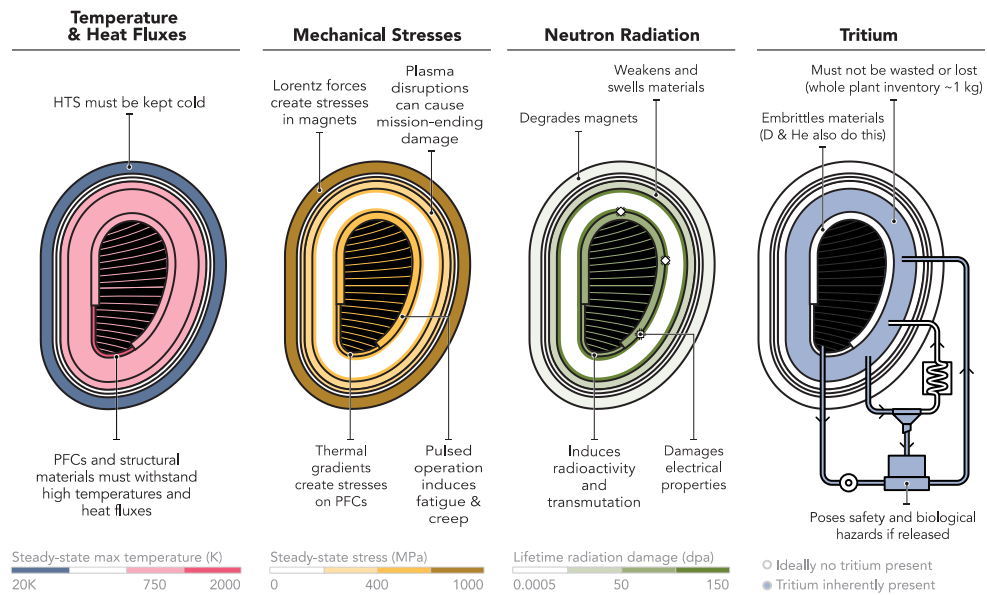

*Figure 2: Schematic distribution of temperature, mechanical stresses, neutron radiation, and tritium in a fusion operating environment. Representative numerical values are shown for Fusion Nuclear Science Facility power plant studies, which feature a Dual Coolant Lead Lithium blanket[18] [19] [20].*

The primary stressors in a fusion environment are shown in Figure 2. They are:

1. *Extreme temperatures and heat fluxes:* Divertors will experience steady-state heat fluxes in the 10 MW/m$^2$ range and correspondingly high transient energy depositions, temperatures, and thermal gradients. First wall, vacuum vessel, blanket will typically operate around 800-1100 K with thermal cycling[18] [21].

2. *Mechanical loads and cycling:* Superconducting magnets must withstand steady-state stresses approaching ~1 GPa. Plasma disruptions can cause maximum stresses as high as 800 MPa in the first wall[22] [23] [18], though these are highly dependent on fusion reactor design and operating mode, emblematic of the high-uncertainty design issues facing fusion power.

3. *Intense radiation fields:* Intense fluxes of 14.1 MeV source neutrons, down-scattered or moderated neutrons, and induced 1–10 MeV gamma-rays cause radiation damage, transmutation, and activation of materials in the core of fusion power plant. Fast neutron fluences on the innermost materials induce 15-50 displacements per atom (dpa) per operational year with 100-1000 atomic parts per million of bulk helium accumulation[24]. Dose rates and energy deposition rates are on par with fission reactors. As despite the 30x higher neutron kinetic energy output per reacting nucleon, DPA rates of fast fission reactors and fusion reactors are predicted to be well within the same order of magnitude.

4. *Plasma-material interactions (PMI):* Materials in the divertor and first wall will experience material erosion and redeposition, material migration and mixing, formation of unique mesostructures such as surface fuzz and gas-filled blisters, and long-term tritium retention, all of which jeopardize the plasma performance and long-term operation of a fusion power plant[25].

5. *Tritium breeding and control:* Materials must be deployed that enable net-tritium breeding[26], control tritium permeation and migration, minimize tritium retention within materials, and keep total onsite tritium inventories in the ~kg per plant range[27] with environmental releases kept as low as possible. A target of <1 g/year is commonly cited [28] but the level that will be tolerated by the public as acceptable may be far lower.

6. *Corrosion.* Structural materials will be exposed to highly corrosive environments, yet the degradation of structural and functional materials must remain low enough to maintain their functionality — or at least degrade little enough — between inspections so as not to require prohibitive or unplanned outages. Experience in the fission energy sector has shown that materials degradation management is one of the main causes of concern for plant outages[29], which themselves drive plant economics, as balancing the short-term needs to shrink outage times lies in tension with longer-term fixes to long-incubation, high-consequence corrosion-based events which can cause fuel failure, prolonged unplanned outages, and in some very real cases, early plant shutdowns. In fusion power plants, coolants and blanket materials will also be subject to intense radiation fluxes and cyclic or transient stresses in addition to corrosion. These effects can couple strongly to each other[30].

## 3. Systems with crucial materials challenges

The materials challenges associated with plasma-facing components (PFCs), blankets, vacuum vessels (VVs), and superconducting magnets have been described in detail in several reviews[31]. No known, currently mass-manufacturable materials can simultaneously withstand the radiation (~$10^{14}$ n/cm$^2$-s), temperature (1100-1500 K) conditions, and crucially the elemental transmutation that will occur in PFCs on the timescale of months or more without critical material property degradation [32,33,34,35,36,37]. In blankets, the breeder material must produce sufficient tritium and lend itself to efficient extraction of both heat and tritium, while the structural materials that surround the blanket must withstand fusion temperatures, neutron fluxes (~0.5-3x$10^{14}$ n/cm$^2$)[38], and must minimize tritium retention[39,40]. VV materials must possess high yield strength, ductility, and thermal conductivity at elevated temperatures, and must retain these properties at high neutron fluences. Moreover, VVs, blankets, and other structural materials will also be exposed to high-temperature working fluids (e.g. coolants such as molten metals like Pb-17Li or molten salts, especially in the case of liquid PFCs[41] or liquid breeder blankets[42]), meaning that corrosion concerns must also be a driving factor in materials selection throughout the fusion power plant[43,44]. Experience from fission power shows us that corrosion, especially insidious, localized modes of corrosion exacerbated by irradiation such as irradiation-assisted stress corrosion cracking (IASCC) of light water reactor (LWR) structural materials or crud-induced localized corrosion (CILC) of Zircaloy fuel cladding, drive pushes for reduced outages and zero fuel failures throughout the fission reactor industry[45,46]. As for electromagnets, their superconducting materials, electrical insulation, and copper must all maintain their performance under radiation despite the presence of a blanket and/or dedicated neutron shielding.

What has received significantly less attention is the fact that as one moves outwards from the plasma, other systems also present substantial materials challenges. These systems have not been a major focus for fusion materials research and development programs until now, and they present many opportunities for researchers traditionally outside of the nuclear materials community to become involved in impactful and necessary work. Fortunately, the further one gets from the plasma, the more similarities exist between fission and fusion, meaning that other fields of materials science are well positioned to contribute to these outstanding challenges:

1. **Plasma heating systems.** Radiofrequency (RF) antennas and waveguides play an essential role in heating the plasma to temperatures that enable fusion. These components must be made of materials that can withstand significant radiation damage and high temperatures, as they must operate directly adjacent to the plasma itself[47]. At present, existing materials and manufacturing processes are not well suited for commercial plants. For example, low-activation steel coated with copper has been considered for RF antennas, but steel has poor thermal conductivity, and plating copper is challenging. Copper alloys such as GRCop-84 have been proposed for RF antenna designs and preliminarily tested in high radiation fields, though integrated antenna performance during or following irradiation has yet to be studied[48]. Moreover, vacuum tube-based RF sources suffer from very short operating lifetimes (a few hours of total use). For example, in a gyrotron (tens of GHz to low THz frequency range), this lifetime is limited by the collector that receives residual power from the electron beam[49]. Meanwhile, solid-state RF sources require new transistor chips with fast dynamics, low ripple factor, and good output regulation[50]. This represents an opportunity for novel chip architectures or substrate materials. The potential for wide-bandgap power-electronic devices for these power systems has also been cited[51]. Finally, ceramics for neutral beam injection insulators and for RF cavity windows are also a challenge because ceramics are generally degraded by irradiation, experiencing radiation-induced electrical conductivity and radiation-induced thermal conductivity degradation[52]. Both of these effects are known to become exacerbated *in operando* during irradiation, yet very little is known about their extent or magnitude across material families.

2. **Heat extraction and power generation.** Commercial-scale fusion power plants require heat exchangers that can operate at very high temperatures, circa 1000 K on the hot side[53]. In several proposed designs, the active material of the breeder blanket (*e.g.*, a molten salt or liquid metal) also acts as the working fluid for the thermal cycle, in which case the heat exchanger materials must be compatible with — and resist corrosion in — this breeder blanket fluid. This is a significant challenge given that many liquid blanket fluids (*e.g.*, FLiBe in the ARC design) are corrosive[39]. Moreover, the working fluid is expected to contain some tritium in practice, so the heat exchanger must resist leakage of tritium while still maximizing heat transfer. The working fluid from the primary coolant will also typically exhibit induced radioactivity from neutron activation, which imposes further demands on the heat exchanger.

3. **Sensors and electronics**. Sensors and electronics for proper measurements and controls will require materials that maintain their electrical performance under neutron irradiation. Among the sensors that will be exposed to the harshest conditions are first wall samples and retroflectors[47]. Ceramics often lose their thermal conductivity and their electrical resistivity under the influence of radiation[52]. Thus, novel types of ceramics are required. This extends beyond sensors and electronics and to several other fusion components (see *Plasma Heating Systems* above). Sensors and electronics tend to be significantly more sensitive to radiation damage than structural alloys. The higher reliance on sensors in fusion compared to fission makes this a critical area of required study.

4. **Fuel cycle.** The primary purpose of the fusion fuel cycle is to replenish tritium fuel via neutron-induced transmutations of Li-containing materials, while minimizing tritium inventory and avoiding tritium releases into the environment. This requires several materials advances for the fuel cycle, including: materials that can efficiently extract tritium from the breeder materials and blanket systems; materials that selectively permeate hydrogen isotopes over helium and other impurities; materials that can separate hydrogen isotopes from each other with high selectivity and high throughput to enable fuel rebalancing; tritium permeation barriers (TPBs) that block the ingress of tritium into coolant channel piping and heat exchanger surfaces; and materials for hydrogen isotope sensing[54]. In liquid breeder concepts, candidate approaches for tritium extraction include electrochemical separation, gas-liquid contactors[55], permeation membranes[56], and vacuum sieve trays[57]. Membranes for metal foil pumps represent one promising pathway for H:He separation to enable direct internal recycling (DIR)[58]. For H isotope separation, porous membranes for temperature-swing adsorption represent a promising direction[59]. A wide range of TPB candidates, including oxides, nitrides, carbon, carbide, MAX-phases and metals have been proposed[60 61]; thus far, it appears that finding a reliable and durable coating process will be at least as challenging at finding a good permeation barrier candidate material.

5. **Maintenance and lifecycle.** Fusion plants will require methods for fully and cost-effectively extracting tritium from solid waste (*e.g.*, irradiated vacuum vessel), liquid waste (*e.g.*, water from detritiation systems), and gaseous waste (*e.g.*, air from HVAC)[62]. There is also a need for radiation-hard components for remote maintenance robots, since conventional microelectronics and robotic systems cannot operate in high radiation environments[63]. For example, in ITER the radiation hardness requirement for robotic handling systems is a total dose of 1 MGy[64].

6. **Insulating components.** Fusion plants will require insulating materials for TBPs, electrical cable interconnects, magnet insulation, and other functions. Ceramics may incur radiation damage and changes to their properties at relatively faster rates than metals, owing to the ionic nature of the bonds, the charge state of some of the defects, and the reduced transport of atoms within the matrix. The resulting changes to properties can be severe — functional ceramics such as $Al_2O_3$ have been shown to gain 5-10 orders of magnitude in electrical conductivity *during* ionizing radiation at fusion-relevant fluxes — limiting their ability to retain electrical resistance when needed in a neutron and/or gamma field[52 65]. In addition, the thermal conductivity of ceramics also rapidly degrades both during and after irradiation. Finally, fusion-relevant fast neutron fluxes can severely degrade the rupture modulus of ceramics[66]. This points to opportunities to both test more functional ceramics in coupled fusion environments and to engineer ceramic compositions more resistant to fusion-relevant neutron and gamma fluxes for operation as the critical components listed above[67].

7. **Joining and welding, especially dissimilar materials.** Even if materials themselves can survive the challenging, coupled environment in a fusion power plant, experience with nuclear fission light water reactors has shown that welds, and the adjacent heat affected zones, often suffer the most severe and catastrophic material failures[68]. Temperature gradients span two orders of magnitude from the PFCs to the magnets. At physical interfaces between components, particularly those that feature large heat fluxes, differences in thermal expansion coefficients make joining dissimilar materials very difficult. Designs that employ molten salts will also likely require corrosion-resistant coatings or claddings that must be sufficiently joined to the vacuum vessel and the breeder blanket. In other industries (*e.g.*, turbine blades), thermal barrier coatings have been engineered to mitigate the effects of thermal gradient induced delamination; however, in the case of fusion, heat must be conducted through the PFCs which complicates the challenge[69]. Proposed solutions include amorphous iron-based phases, creating material gradients between tungsten and RAFM steels, functional gradient deposition, or diffusion bonding using intermetallic[70 71 72 73 74]. These approaches have yet to be tested under high heat fluxes or high neutron irradiation to mimic fusion plant environments.

8. **Materials testing and scalable manufacturing.** It is well established that developing a novel material in the laboratory is only the first step. This calls for the need for an end-to-end pipeline from materials discovery to scaleup. While facilities like the Integrated Fusion Material Testing Irradiation Facility (IFMIF) and the Fusion Prototypic Neutron Source (FPNS) will represent the last, crucial stage of material testing before selecting final materials for a fusion power plant, a plethora of mid-scale coupled effects facilities — such as high-flux proton irradiation stations, high heat flux test facilities, coupled corrosion-radiation-magnetic-field-stress test facilities, and other such coupled effects facilities — must be commissioned as gates through which to down-select only the most promising material candidates for IFMIF/FPNS neutron irradiation testing. Commercial scalability requires significant advances in manufacturability, a robust and reliable supply chain, as well as material certification and qualification.

The above is not a comprehensive list (we did not discuss materials opportunities in neutron shielding and neutron multipliers, for example), but it represents an important illustrative view of historically less-funded materials challenges for fusion energy. Ultimately, the evolution from early devices made from available materials to commercially scalable ones made from new, more suitable materials is not uncommon: a possible trajectory for the fusion energy sector might be compared with the trajectory of the civilian aviation in the 20$^{th}$ century (Table 1).

*Table 1: Evolution from proof of concept to market transformation in the civilian aviation and fusion energy sectors. Project names from Commonwealth Fusion Systems (CFS) (e.g. SPARC, ARC) are given for illustrative purposes. Materials researchers today can primarily impact future generations of fusion power plants ("Commercially Viable" and "Market Transformation"), although applied materials engineering is still highly relevant to pilot and first-of-a-kind devices in the near term.*

|  | **Enabling Component** | **Proof of Concept Device** | **Pilot and First of a Kind** | **Commercially Viable** | **Market Transformation** |
|---|---|---|---|---|---|
| **Civilian Aviation** | High aspect ratio wings (1900s) *Wood & fabric* | Wright Flyer (1900s) *Wood & fabric* | Fokker (1910s) *Steel* | Boeing 707 (1950s) *Aluminum* | Boeing 747 (1970s) *Aluminum + composites* |
| **Fusion Energy** (example: CFS) | SPARC toroidal field model coil (2021) | SPARC (Stated target: 2020s) | First few ARC plants (Stated target: 2030s) | First fleet of ARC plants with more advanced materials & systems (2040s?) | Fusion as a meaningful % of global energy (2060s?) |

## 4. Pathways for Fusion-Materials Innovation in a High-Uncertainty Environment

**Design uncertainty in fusion.** Thus far, we have outlined some of the major opportunities for materials research in the fusion energy field. Fusion is qualitatively different than many of the research areas that we as materials scientists are accustomed to. On a spectrum from low design uncertainty (*e.g.*, single-junction solar cells) to high design uncertainty materials problems (*e.g.*, materials for quantum computers), fusion energy sits on the high side. We discuss the implications for materials scientists of the high design uncertainty that characterizes fusion R&D, and we present a framework that can orient the materials science community towards having a maximum impact.

Taken as a whole, the materials science community is well accustomed to "discovery-oriented" and to "optimization-oriented" research[75]; fusion fits into neither category. Discovery-oriented materials research seeks to explore in an open-ended manner; this will not do for fusion, because there is a specific end goal in mind (practical fusion energy on a decadal timeframe). In contrast, optimization-oriented research seeks to meet or exceed fixed target properties. Silicon photovoltaics and lithium-ion batteries over the past decade are good examples: materials scientists had a clear set of performance and cost targets to design for, a finite set of market applications, a 'dominant design,' and quantifiable costs of production to feed into technoeconomic models[76][77][78]. For fusion, the design landscape is far too uncharted — and the need for innovation too pressing — for conventional optimization: many of the materials-property targets depend on the final reactor design, which is not yet known; the latter in turn depends on what is achievable materials-wise. As a result, fusion energy involves high design uncertainty (see Supplemental Information for a categorization of types of uncertainty in materials for fusion).

How, then, are materials researchers to develop and optimize materials in this context, when there is no good way to test candidate materials in a representative fusion environment, when each fusion project employs a different reactor design, and when every materials choice must take into consideration other neighboring systems, which have yet to be finalized? These sources of uncertainty can seem paralyzing.

It may be tempting to wait it out, *i.e.*, to take a "serial" innovation approach in which materials innovation only begins *after* a system design freeze. However, this is not a realistic option. System designers, plasma engineers, government regulators, companies, and other stakeholders cannot finalize a system without frequent input from the materials science community. Furthermore, climate change and other pressing environmental challenges cannot wait for serial innovation to take its time. A different approach is needed.

Some high-level recommendations arise from the characteristics of the fusion field today. Given the relative uncertainty in quantitative performance requirements for many fusion materials, it is ineffective to try to optimize for a fixed objective; rather, many R&D efforts should focus on mapping out the space of what is possible. For example, whereas some materials development programs are aimed at developing a single optimal material for a given use case, the fusion materials community should concentrate instead on producing high-throughput fixtures that afford high sample quality and reproducibility: the emphasis should be on exploration and the ability to iterate.

Moreover, while it is certainly the case that materials testing facilities are sorely needed, it would be unwise to wait for a fully representative test environment (e.g., a Fusion Prototypic Neutron Source or a national tritium laboratory); instead, the community can start using imperfect tools as heuristics, such as simulations, ion sources as proxy for neutron sources, deuterium as proxy for tritium, and so forth. Given the tight interdependency between materials and power plant designs, successful materials development efforts should build venues for rapid feedback about cost, system requirements, environmental considerations, integrations with other subsystems, etc. In other words, materials performance should not be the sole consideration in these development efforts.

Finally, it will be essential that the fusion community converges more than it diverges, as opposed to constantly proposing new materials and new fusion designs but rarely discarding any avenues of inquiry. The way to ensure convergence is to build in processes and frameworks for rapidly evaluating reactor-material combinations. Based on this overall R&D philosophy, we propose an approach to fusion materials development inspired by the iterative innovation model[79] and co-evolution design[80] field, as well as from the natural and structured development styles characterized by Looney et al. [81], and we articulate how materials researchers can employ this approach to maximize their impact in the field of fusion energy.

**Iterative co-evolution.** In the proposed research framework, materials development efforts and system design targets co-evolve and, over the course of multiple iterations, converge together on a dominant design (Figure 3). The goal is to converge on a combination of viable materials and a viable fusion plant design that are compatible with each other. The materials science & fusion plant development communities start with their respective list of candidate materials and plant designs, and they strive to rule out infeasible options. This requires active collaborations and dialogue between these two communities. In the process, new candidates emerge through research and are evaluated alongside existing candidates.

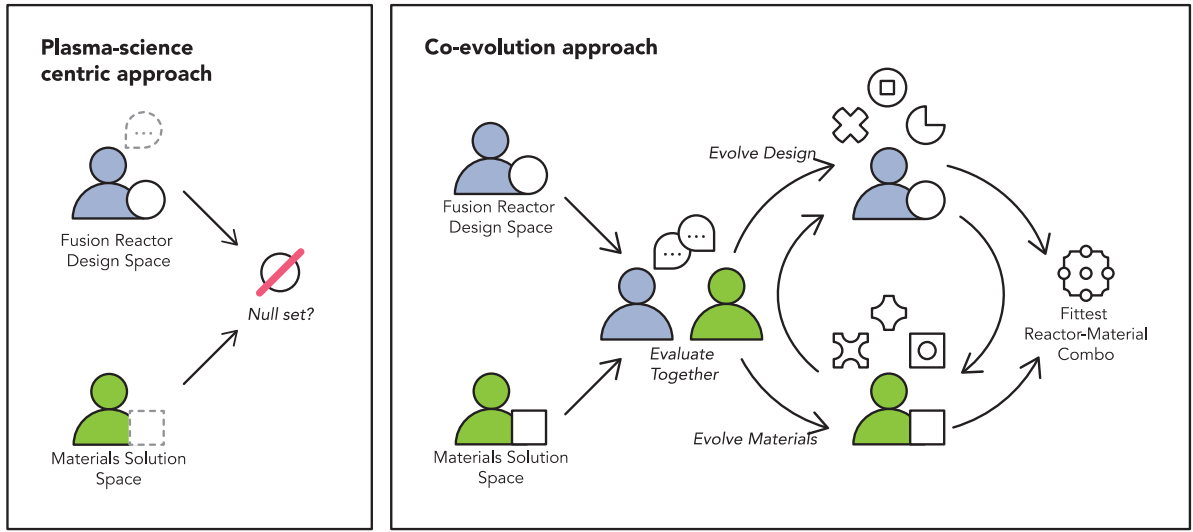

*Figure 3: Iterative co-evolution approach to materials science for fusion. The fusion power plant design community (e.g., commercial fusion companies) and the materials science community periodically come together to share their latest candidates with each other, and then go back to keep refining candidates based on what they learned during these exchanges. This iterative process continues until there is a set of materials and a plant design that are compatible with each other.*

In this process, the act of excluding candidate materials is as valuable as generating them. Rather than striving to select 'winners' among candidate materials, independent materials science groups must instead prioritize the ruling out of candidates that no longer appear promising for any fusion plant design. For research labs studying materials, this might take the form of excluding tungsten as a first wall candidate beyond the pilot plant stage because of its propensity for neutron embrittlement, and of focusing instead on capacity-building for substitute materials with improved properties. This iterative approach allows materials scientists to conduct meaningful research that advances the fusion field without waiting for a final plant design; conversely, it allows plant developers to design prototype and pilot devices without waiting for the perfect fusion materials. Naturally, the act of excluding candidate materials goes against the well-documented human tendency to keep investing in efforts that one has already invested in, even when they cease to look promising. This requires built-in mechanisms to avoid this sunk cost fallacy effect, both at the individual researcher level (e.g., self-awareness) and at the organizational or R&D community level (e.g., more frequent project performance evaluations)[82].

The co-evolution design framework may feel unfamiliar to researchers coming from other areas of materials science, but many elements of this design philosophy are already well accepted in the fusion materials field. For example, key aspects of co-evolution design can already be found in the R&D approaches of the DEMO concept developed by the EUROfusion consortium and the STEP concept developed by the UK Atomic Energy Authority[83][84]. UK-STEP and EU-DEMO employ a concurrent design approach, whereby multiple sub-systems are designed concurrently with a common simulation framework to define touchpoints and handoffs between them. These devices are designed such that components can be replaced — and new materials can be tested — in the device over time, so that learnings from materials research can be quickly implemented.

**Capabilities needed.** Multiple capabilities must be leveraged to enable the iterative co-design process illustrated in Figure 3. Many of these capabilities exist in the broader materials research community, providing opportunities to expand the fusion workforce via cross-cutting collaborations. Successful fusion materials discovery will require contributions from experimental materials, fundamental materials modeling, scientific machine learning, and agile project management. To enable the iterative co-design process illustrated in Figure 3, several key capabilities are needed:

- *The ability to 'handshake' between communities:* The materials science and power plant design communities need avenues to exchange information. This makes it possible to evaluate materials options in the context of specific plant design options, and to discard materials that are not compatible with any plant design options (and conversely, to discard plant design options that are not feasible with any envisioned materials).

- *The ability to discard candidates*, *e.g.*, to stop researching materials that cannot withstand the expected combined-stressor environments, or whose cost even at scale would still be uneconomical. This requires a forum for the materials science community to compare results and debate which options should be abandoned for the time being, and under which conditions these options should be revisited. It also requires technoeconomic analysis capabilities to rapidly estimate the cost, scalability, and supply-chain considerations of candidate materials. These capabilities need not be complex: simple technoeconomic models that rely on parametric relationships and heuristics may be sufficient in some cases. It is essential that these decision-making processes be able to respond quickly to changing supply-chain realities, yet have slightly dampened response functions at early technology readiness levels (TRLs) to allow broader exploration. Care and community consensus must be frequently sought to move the field forward, sometimes at the expense of the preferences of individual research groups that have built a specialization in materials systems that are no longer relevant to fusion power plants.

- *The ability to pivot:* Rapid and frequent switching is an inherent part of the iterative process. For materials scientists, this means investing in materials explorations platforms that are versatile, *i.e.,* can perform assessments on wide ranges of materials. This makes it possible to facilities and labs to remain at the forefront even as materials candidates are discarded. At the same time, given changing plant designs, regulatory considerations, or supply chain realities, it is ideal to maintain latent capability to revisit previously discarded materials, as they may be reactivated later.

- *The ability to conceive of new candidate materials, and to prioritize which candidates to synthesize and test* (especially given the difficulty of testing materials for fusion). Two proposed approaches include first-principles simulations accelerated using machine learning, and/or computationally informed conditional generative design (inverse design). Currently these tools can predict limited numbers of materials for "hero experiments,"[85] [86] and have yet to generalize across broader ranges of materials, including composites, alloys, and microstructure-engineered materials. Additionally, to identify truly exceptional materials, these models must accurately propose candidates well outside of their training-set distributions[87].

- *The ability to rapidly synthesize and test:* Experimental validation of predicted compounds requires synthesis tools to be (i) flexible: synthesize a wide range of compounds with a wide range of processing conditions (usually ~10 unique processing attempts per given composition[88]); (ii) precise: synthesize high-quality materials with precise elemental ratios, morphology, and crystallographic structure; and (iii) fast: estimated throughputs of 100–1000 samples per minute are required to canvas materials space given the accuracy of today's generative design algorithms (which may improve with time). Currently, few laboratories can achieve these throughputs with high quality and reproducibility. Self-driving labs, in which synthesis and analysis occur in closed-loop fashion, combining automation and machine learning, represent a promising path. We are also optimistic about the role of high-throughput materials synthesis techniques[89] and advanced sample production techniques and additive manufacturing of fusion-relevant materials[90] [91]. For the synthesis step specifically, the community bifurcates between scaling up traditional synthesis techniques (*e.g.*, ball milling and powder sintering) and harnessing the power of miniaturization (*e.g.*, printing, additive manufacturing, and micro-sputtering in combinatorial fashion). The net result is to minimize testing time (or time-in-test) per material. Some of us have argued elsewhere that finding exceptional materials requires not only more attempts per unit time (sample throughput) but also higher success probability (predictive accuracy)[92]. The discovery of exceptional materials may involve experimental screening of highly non-convex, non-smooth, discontinuous search spaces, such as the search for "islands of opportunity" as materials whose properties do not follow as linear combinations of their constituents, and the development of tools to navigate such spaces in a resource-efficient manner.[87] As discussed above, testing facilities for fusion materials are scarce; there is a need for new facilities to expose candidate materials to relevant temperatures, mechanical loadings, neutron irradiation, tritium concentrations, and so forth[18], and some or all of these effects at the same time.

- *The ability to gain deeper materials understanding through simulation*: Computational modeling can complement or even preempt experimental testing given the difficulty of performing realistic radiation testing[93], for example by helping to predict macroscopic mechanical evolution and failure onset during operation[38]. Recent advances in capabilities for atomistic simulations enable much more predictive simulations than was possible even a decade ago. We see an opportunity for simulation approaches that can link atomistic effects with mesoscale properties, for example to predict macroscopic mechanical evolution and failure onset during operation[94]. However, the persistent gap between simulation and experiment in materials research implies that modeling should still be connected to experiment. Ideally, pathways are created to link experimental results back to refine and improve modeling. In particular, machine learning surrogate models combined with scientific machine learning (to extract learned equations from trained models) may present opportunities.

- *The ability to develop robust material property inference models*: Oftentimes the properties of ultimate interest, such as long-term corrosion rates, creep performance, or cyclic fatigue lifetime, are prohibitively difficult or time-consuming to enable rapid materials innovation. This opens up opportunities to establish strong correlations between more readily measured material properties and those of ultimate interest. Perfect matches or causal relationships may not be required in all cases. For example, the more readily measured hardness has been linked to yield strength in some material families[95], and void swelling during irradiation can be detected via changes in porosity-affected sound speed[96]. Such inference models, rooted in faster *in situ* experiments, enable both the application of more coupled environmental effects and more rapid data generation to more quickly down-select the most viable candidate materials for further testing.

- *The ability to test in realistic operating environments*, both *in situ* and *in operando*. This requires investing in test facilities, including combined-effects testing facilities[97]. In adjacent fields, two kinds of tests have proven valuable in general: (1) ones that can rapidly produce directionally correct results, so called "proxy measurements," which can be cheaper and safer to run than high-fidelity tests, and (2) ones that can replicate coupled effects, e.g., the combined effect of high heat fluxes and neutron radiation. This may well require building specialized facilities that can perform materials testing as a service, as well as smaller proxy testing tools within individual laboratories that can provide directional feedback on a shorter timeframe.

- *The ability to employ Artificial Intelligence (AI) tools that incorporate complexity and uncertainty*. The sub-fields of AI that pertain to *s*cientific machine learning (ML), uncertainty estimation, route planning, and generative design may find broad applications in materials and plant design[98][99][100]. Specific areas deemed ripe for application of these methods include materials down-selection and design, materials-synthesis hardware control, optimization of material composition and, accelerated characterization, durability testing with combined stressors, root-cause analysis, and extraction of scientific equations[101].

In summary, the materials science community should focus on capability building and on developing avenues for rapid iterative feedback with the fusion power plant community, and on obtaining high-quality information with low switching cost.

## 5. Conclusion

We have presented some key opportunities for materials scientists to make high-impact contributions to the fusion field, and described the ways in which high design uncertainty calls for a fundamentally different approach to materials research. This article is intentionally regional in perspective and aims to be illustrative rather than comprehensive, but the general principles should apply more broadly in the fusion field. It is meant to catalyze conversation and share pioneer learnings, not to replace a comprehensive community roadmap.

Our perspective makes clear the importance of accelerating the feedback loops between technology (materials researchers, plasma physicists), market (power plant designers, investors, technoeconomic analysis experts), and governing bodies (regulators, government agencies). These feedback loops could take many forms: regular in-person workshops, community roadmap reports, online project status databases, interdisciplinary project teams, and so forth. Ultimately, the development of materials for commercial fusion energy will require broad participation of the materials science & engineering community. This is the time to start ramping these research efforts up if fusion is to play a meaningful role in the clean energy mix of the 21$^{st}$ century.

**Acknowledgments**

The authors thank Marta Barriga for upgrading our figure illustrations; Rodrigo Freitas, Theodore Mouratidis, and Kevin Woller for providing ideas and feedback; Warren Seering and Gene Fitzgerald for discussions about co-evolution design and iterative innovation, respectively; and MIT Proto Ventures and Parviz Tayebati for useful discussions and for support.

**Author contributions**

Conceptualization: D.C.T., M.P.S., S.E.F., and T.B.; Methodology: D.C.T. and T.B.; Investigation, D.C.T.: M.P.S., M.G.S, M.P.S., S.E.F., Z.S.H., and Z.S.H.; Writing - Original Draft: D.C.T., M.P.S., and T.B.; Writing - Reviewing & Editing: M.G.S, D.G.W., M.P.S., S.E.F., T.B., and Z.S.H.; Visualization: D.C.T., M.G.S, and T.B. ; Supervision: , D.G.W. and T.B.; Project Administration: D.C.T.; Funding Acquisition: D.G.W. and T.B.

**Declaration of interests**

The authors declare no competing interests.

**Bibliography**

1. Schwartz, J.A., Ricks, W., Kolemen, E., and Jenkins, J.D. (2023). The value of fusion energy to a decarbonized United States electric grid. Joule *7*, 675–699. https://doi.org/10.1016/j.joule.2023.02.006.

2. With historic explosion, a long sought fusion breakthrough https://www.science.org/content/article/historic-explosion-long-sought-fusion-breakthrough.

3. Ciattaglia, S., Federici, G., Barucca, L., Lampasi, A., Minucci, S., and Moscato, I. (2017). The European DEMO fusion reactor: Design status and challenges from balance of plant point of view. In 2017 IEEE International Conference on Environment and Electrical Engineering and 2017 IEEE Industrial and Commercial Power Systems Europe (EEEIC / I&CPS Europe), pp. 1–6. https://doi.org/10.1109/EEEIC.2017.7977853.

4. Wan, Y., Li, J., Liu, Y., Wang, X., Chan, V., Chen, C., Duan, X., Fu, P., Gao, X., Feng, K., et al. (2017). Overview of the present progress and activities on the CFETR. Nucl. Fusion *57*, 102009. https://doi.org/10.1088/1741-4326/aa686a.

5. Anand, H., Bardsley, O., Humphreys, D., Lennholm, M., Welander, A., Xing, Z., Barr, J., Walker, M., Mitchell, J., and Meyer, H. (2023). Modelling, design and simulation of plasma magnetic control for the Spherical Tokamak for Energy Production (STEP). Fusion Eng. Des. *194*, 113724. https://doi.org/10.1016/j.fusengdes.2023.113724.

6. Creely, A.J., Greenwald, M.J., Ballinger, S.B., Brunner, D., Canik, J., Doody, J., Fülöp, T., Garnier, D.T., Granetz, R., Gray, T.K., et al. (2020). Overview of the SPARC tokamak. J. Plasma Phys. *86*, 865860502. https://doi.org/10.1017/S0022377820001257.

7. Fusion Industry Association (2023). The global fusion industry in 2023.

8. Parisi, J., and Ball, J. (2018). The Future of Fusion Energy (WSPC).

9. Freidberg, J.P. (2007). Plasma Physics and Fusion Energy (Cambridge University Press) https://doi.org/10.1017/CBO9780511755705.

10. Fusion Energy Sciences Advisory Committee (2018). Transformative Enabling Capabilities for Efficient Advance Toward Fusion Energy (U.S Department of Energy).

11. Carter, T., Baalrud, S., Betti, R., Ellis, T., Foster, J., Geddes, C., Gleason, A., Holland, C., Humrickhouse, P., Kessel, C., et al. (2020). Powering the Future: Fusion & Plasmas (US Department of Energy (USDOE), Washington, DC (United States). Office of Science) https://doi.org/10.2172/1995209.

12. Baalrud, S., Ferraro, N., Garrison, L., Howard, N., Kuranz, C., Sarff, J., Scime, E., Solomon, W., Biewer, T., Brunner, D., et al. A Community Plan for Fusion Energy and Discovery Plasma Sciences.

13. Bringing Fusion to the U.S. Grid (2021). (National Academies Press) https://doi.org/10.17226/25991.

14. Zinkle, S.J., and Snead, L.L. (2014). Designing Radiation Resistance in Materials for Fusion Energy. Annu. Rev. Mater. Res. *44*, 241–267. https://doi.org/10.1146/annurev-matsci-070813-113627.

15. Knaster, J., Moeslang, A., and Muroga, T. (2016). Materials research for fusion. Nat. Phys. *12*, 424–434. https://doi.org/10.1038/nphys3735.

16. Was, G.S., Petti, D., Ukai, S., and Zinkle, S. (2019). Materials for future nuclear energy systems. J. Nucl. Mater. *527*, 151837. https://doi.org/10.1016/j.jnucmat.2019.151837.

17. Gilbert, M.R., Arakawa, K., Bergstrom, Z., Caturla, M.J., Dudarev, S.L., Gao, F., Goryaeva, A.M., Hu, S.Y., Hu, X., Kurtz, R.J., et al. (2021). Perspectives on multiscale modelling and experiments to accelerate materials development for fusion. J. Nucl. Mater. *554*, 153113. https://doi.org/10.1016/j.jnucmat.2021.153113.

18. Rowcliffe, A.F., Garrison, L.M., Yamamoto, Y., Tan, L., and Katoh, Y. (2017). Materials challenges for the fusion nuclear science facility. Fusion Eng. Des. *135*. https://doi.org/10.1016/j.fusengdes.2017.07.012.

19. Huang, Y., Tillack, M.S., Ghoniem, N.M., Blanchard, J.P., El-Guebaly, L.A., and Kessel, C.E. (2018). Multiphysics modeling of the FW/Blanket of the U.S. fusion nuclear science facility (FNSF). Fusion Eng. Des. *135*, 279–289. https://doi.org/10.1016/j.fusengdes.2017.07.005.

20. Humrickhouse, P.W., and Merrill, B.J. (2018). Tritium aspects of the fusion nuclear science facility. Fusion Eng. Des. *135*, 302–313. https://doi.org/10.1016/j.fusengdes.2017.04.099.

21. Pintsuk, G., Aiello, G., Dudarev, S.L., Gorley, M., Henry, J., Richou, M., Rieth, M., Terentyev, D., and Vila, R. (2022). Materials for in-vessel components. Fusion Eng. Des. *174*, 112994. https://doi.org/10.1016/j.fusengdes.2021.112994.

22. Sugihara, M., Shimada, M., Fujieda, H., Gribov, Y., Ioki, K., Kawano, Y., Khayrutdinov, R., Lukash, V., and Ohmori, J. (2007). Disruption scenarios, their mitigation and operation window in ITER. Nucl. Fusion *47*, 337–352. https://doi.org/10.1088/0029-5515/47/4/012.

23. Kwon, S., and Hong, S.-H. (2021). Estimation of electromagnetic loads including eddy and halo current on the major disruption for the K-DEMO divertor. Fusion Eng. Des. *168*, 112592. https://doi.org/10.1016/j.fusengdes.2021.112592.

24. Bocci, B., Hartwig, Z., Segantin, S., Testoni, R., Whyte, D., and Zucchetti, M. ARC reactor materials: activation analysis and optimization.

25. Federici, G., Skinner, C.H., Brooks, J.N., Coad, J.P., Grisolia, C., Haasz, A.A., Hassanein, A., Philipps, V., Pitcher, C.S., Roth, J., et al. (2001). Plasma–material interactions in current tokamaks and their implications for next step fusion reactors. Nucl. Fusion *41*.

26. Meschini, S., Ferry, S.E., Delaporte-Mathurin, R., and Whyte, D.G. (2023). Modeling and analysis of the tritium fuel cycle for ARC- and STEP-class D-T fusion power plants. Nucl. Fusion *63*, 126005. https://doi.org/10.1088/1741-4326/acf3fc.

27. Abdou, M., Riva, M., Ying, A., Day, C., Loarte, A., Baylor, L.R., Humrickhouse, P., Fuerst, T.F., and Cho, S. (2020). Physics and technology considerations for the deuterium–tritium fuel cycle and conditions for tritium fuel self sufficiency. Nucl. Fusion *61*, 013001. https://doi.org/10.1088/1741-4326/abbf35.

28. Larsen, G., and Babineau, D. (2020). An Evaluation of the Global Effects of Tritium Emissions from Nuclear Fusion Power. Fusion Eng. Des. *158*, 111690. https://doi.org/10.1016/j.fusengdes.2020.111690.

29. Busby, J.T., Nanstad, R.K., Stoller, R.E., Feng, Z., and Naus, D.J. (2008). Materials Degradation in Light Water Reactors: Life After 60 https://doi.org/10.2172/938766.

30. Zhou, W., Yang, Y., Zheng, G., Woller, K.B., Stahle, P.W., Minor, A.M., and Short, M.P. (2020). Proton irradiation-decelerated intergranular corrosion of Ni-Cr alloys in molten salt. Nat. Commun. *11*, 3430. https://doi.org/10.1038/s41467-020-17244-y.

31. Zinkle, S.J., and Quadling, A. (2022). Extreme materials environment of the fusion "fireplace." MRS Bull. *47*, 1113–1119. https://doi.org/10.1557/s43577-022-00453-9.

32. Woller, K.B., Whyte, D.G., and Wright, G.M. (2017). Impact of helium ion energy modulation on tungsten surface morphology and nano-tendril growth. Nucl. Fusion *57*, 066005. https://doi.org/10.1088/1741-4326/aa67ac.

33. Litnovsky, A., Schmitz, J., Klein, F., De Lannoye, K., Weckauf, S., Kreter, A., Rasinski, M., Coenen, J.W., Linsmeier, C., Gonzalez-Julian, J., et al. (2020). Smart Tungsten-based Alloys for a First Wall of DEMO. Fusion Eng. Des. *159*, 111742. https://doi.org/10.1016/j.fusengdes.2020.111742.

34. Smolentsev, S., Rognlien, T., Tillack, M., Waganer, L., and Kessel, C. (2019). Integrated Liquid Metal Flowing First Wall and Open-Surface Divertor for Fusion Nuclear Science Facility: Concept, Design, and Analysis. Fusion Sci. Technol. *75*, 939–958. https://doi.org/10.1080/15361055.2019.1610649.

35. Wang, X., Huang, H., Shi, J., Xu, H.-Y., and Meng, D.-Q. (2021). Recent progress of tungsten-based high-entropy alloys in nuclear fusion. Tungsten *3*, 143–160. https://doi.org/10.1007/s42864-021-00092-8.

36. Barron, P. (2021). The Development of Low-Activation, Multi-Principal Element Alloys for Nuclear Fusion Applications.

37. Neuman, E.W., Hilmas, G.E., and Fahrenholtz, W.G. (2016). Ultra-High Temperature Mechanical Properties of a Zirconium Diboride–Zirconium Carbide Ceramic. J. Am. Ceram. Soc. *99*, 597–603. https://doi.org/10.1111/jace.13990.

38. UK Atomic Energy Authority, and Henry Royce Institute (2021). UK Fusion Materials Roadmap.

39. Patel, N.S., Pavlík, V., and Boča, M. (2017). High-Temperature Corrosion Behavior of Superalloys in Molten Salts – A Review. Crit. Rev. Solid State Mater. Sci. *42*, 83–97. https://doi.org/10.1080/10408436.2016.1243090.

40. Zhou, W., Yang, Y., Zheng, G., Woller, K.B., Stahle, P.W., Minor, A.M., and Short, M.P. (2020). Proton irradiation-decelerated intergranular corrosion of Ni-Cr alloys in molten salt. Nat. Commun. *11*, 3430. https://doi.org/10.1038/s41467-020-17244-y.

41. Andruczyk, D., Maingi, R., Kessel, C., Curreli, D., Kolemen, E., Canik, J., Pint, B., Youchison, D., and Smolentsev, S. (2020). A Domestic Program for Liquid Metal PFC Research in Fusion. J. Fusion Energy *39*, 441–447. https://doi.org/10.1007/s10894-020-00259-0.

42. Sorbom, B., Ball, J., Barnard, H., Haakonsen, C., Hartwig, Z., Olynyk, G., Sierchio, J., and Whyte, D. (2012). Liquid immersion blanket design for use in a compact modular fusion reactor. In APS Division of Plasma Physics Meeting Abstracts, pp. UP8-055.

43. Gnanasekaran, T., Dayal, R.K., and Raj, B. (2012). 10 Liquid metal corrosion in nuclear reactor and accelerator driven systems. In Nuclear Corrosion Science and Engineering, pp. 301–328. https://doi.org/10.1533/9780857095343.3.301.

44. Zheng, G., and Sridharan, K. (2018). Corrosion of Structural Alloys in High-Temperature Molten Fluoride Salts for Applications in Molten Salt Reactors. JOM *70*, 1535–1541. https://doi.org/10.1007/s11837-018-2981-2.

45. Was, G.S., Bahn, C.-B., Busby, J., Cui, B., Farkas, D., Gussev, M., He, M.R., Hesterberg, J., Jiao, Z., and Johnson, D. (2024). How irradiation promotes intergranular stress corrosion crack initiation. Prog. Mater. Sci., 101255.

46. Deshon, J., Hussey, D., Kendrick, B., McGurk, J., Secker, J., and Short, M. (2011). Pressurized water reactor fuel crud and corrosion modeling. JOM *63*, 64–72. https://doi.org/10.1007/s11837-011-0141-z.

47. Alba, R., Iglesias, R., and Cerdeira, M.Á. (2022). Materials to Be Used in Future Magnetic Confinement Fusion Reactors: A Review. Materials *15*, 6591. https://doi.org/10.3390/ma15196591.

48. Seltzman, A.H., and Wukitch, S.J. (2020). Nuclear response of additive manufactured GRCop-84 copper for use in Lower hybrid launchers in a fusion environment. Fusion Eng. Des. *159*, 111726. https://doi.org/10.1016/j.fusengdes.2020.111726.

49. Ives, R.L., Singh, A., and Borchard, P. (2007). Improved collectors for high power gyrotrons. In 2007 Joint 32nd International Conference on Infrared and Millimeter Waves and the 15th International Conference on Terahertz Electronics, pp. 436–437. https://doi.org/10.1109/ICIMW.2007.4516569.

50. Patel, P., Sumod, C.B., Thakkar, D.P., Gupta, L.N., Patel, V.B., Bansal, L.K., Qureshi, K., Vadher, V., Singh, N.P., and Baruah, U.K. (2013). Development Overview of Solid-State Multimegawatt Regulated High-Voltage Power Supplies Utilized by NBI and RF Heating Systems. IEEE Trans. Plasma Sci. *41*, 263–268. https://doi.org/10.1109/TPS.2012.2227330.

51. Scott, M.J., Wang, J., Fu, P., Yang, L., Gao, G., Sheng, Z., Huang, L., and Yuan, H. (2015). Wide bandgap switching devices for fusion reactor power supply systems. In 2015 IEEE 26th Symposium on Fusion Engineering (SOFE), pp. 1–6. https://doi.org/10.1109/SOFE.2015.7482354.

52. Zinkle, S.J., and Hodgson, E.R. (1992). Radiation-induced changes in the physical properties of ceramic materials. J. Nucl. Mater. *191–194*, 58–66. https://doi.org/10.1016/S0022-3115(09)80011-1.

53. Segantin, S., Bersano, A., Falcone, N., and Testoni, R. (2020). Exploration of power conversion thermodynamic cycles for ARC fusion reactor. Fusion Eng. Des. *155*, 111645. https://doi.org/10.1016/j.fusengdes.2020.111645.

54. Taylor, C.N. (2022). Hydrogen and its detection in fusion and fission nuclear materials – a review. J. Nucl. Mater. *558*, 153396. https://doi.org/10.1016/j.jnucmat.2021.153396.

55. Alpy, N., Terlain, A., and Lorentz, V. (2000). Hydrogen extraction from Pb–17Li: results with a 800 mm high packed column. Fusion Eng. Des. *49–50*, 775–780. https://doi.org/10.1016/S0920-3796(00)00461-0.

56. Humrickhouse, P.W., and Merrill, B.J. (2015). Vacuum Permeator Analysis for Extraction of Tritium from DCLL Blankets. Fusion Sci. Technol. *68*, 295–302. https://doi.org/10.13182/FST14-941.

57. Mertens, M.A.J., Demange, D., and Frances, L. (2016). Model and simulation of a vacuum sieve tray for T extraction from liquid PbLi breeding blankets. Fusion Eng. Des. *112*, 541–547. https://doi.org/10.1016/j.fusengdes.2016.05.038.

58. Haertl, T., Day, C., Giegerich, T., Hanke, S., Hauer, V., Kathage, Y., Lilburne, J., Morris, W., and Tosti, S. (2022). Design and feasibility of a pumping concept based on tritium direct recycling. Fusion Eng. Des. *174*, 112969. https://doi.org/10.1016/j.fusengdes.2021.112969.

59. Neugebauer, C., Hörstensmeyer, Y., and Day, C. (2020). Technology Development for Isotope Rebalancing and Protium Removal in the EU-DEMO Fuel Cycle. Fusion Sci. Technol. *76*, 215–220. https://doi.org/10.1080/15361055.2019.1704139.

60. Rönnebro, E.C.E., Oelrich, R.L., and Gates, R.O. (2022). Recent Advances and Prospects in Design of Hydrogen Permeation Barrier Materials for Energy Applications—A Review. Molecules *27*, 6528. https://doi.org/10.3390/molecules27196528.

61. Causey, R.A., Karnesky, R.A., and Marchi, C.S. (2012). Tritium Barriers and Tritium Diffusion in Fusion Reactors. Compr. Nucl. Mater. *4*, 511–549. https://doi.org/10.1016/b978-0-08-056033-5.00116-6.

62. Reynolds, S., Newman, M., Coombs, D., and Witts, D. (2016). JET experience on managing radioactive waste and implications for ITER. Fusion Eng. Des. *109–111*, 979–985. https://doi.org/10.1016/j.fusengdes.2016.01.039.

63. Crofts, O., Loving, A., Torrance, M., Budden, S., Drumm, B., Tremethick, T., Chauvin, D., Siuko, M., Brace, W., Milushev, V., et al. (2022). EU DEMO Remote Maintenance System development during the Pre-Concept Design Phase. Fusion Eng. Des. *179*, 113121. https://doi.org/10.1016/j.fusengdes.2022.113121.

64. Saito, M., Anzai, K., Maruyama, T., Noguchi, Y., Ueno, K., Takeda, N., and Kakudate, S. (2016). Development of radiation hard components for ITER blanket remote handling system. Fusion Eng. Des. *109–111*, 1502–1506. https://doi.org/10.1016/j.fusengdes.2015.11.042.

65. White, D.P., Snead, L.L., Zinkle, S.J., and Eatherly, W.S. (1998). In situ measurement of radiation induced conductivity in oxide insulators during neutron irradiation. J. Appl. Phys. *83*, 1924–1930. https://doi.org/10.1063/1.366917.

66. Katoh, Y., Snead, L.L., Parish, C.M., and Hinoki, T. (2013). Observation and possible mechanism of irradiation induced creep in ceramics. J. Nucl. Mater. *434*, 141–151. https://doi.org/10.1016/j.jnucmat.2012.11.035.

67. Hopkins, G.R., and Price, R.J. (1985). Fusion reactor design with ceramics. Nucl. Eng. Des. Fusion *2*, 111–143. https://doi.org/10.1016/0167-899X(85)90008-4.

68. Canonico, D.A. (1977). Significance of reheat cracks to the integrity of pressure vessels for light-water reactors (Oak Ridge National Lab. (ORNL), Oak Ridge, TN (United States)) https://doi.org/10.2172/7101283.

69. Bumgardner, C., Croom, B., and Li, X. (2017). High-temperature delamination mechanisms of thermal barrier coatings: In-situ digital image correlation and finite element analyses. Acta Mater. *128*, 54–63. https://doi.org/10.1016/j.actamat.2017.01.061.

70. Wu, X., Kondo, S., Yu, H., Okuno, Y., Ando, M., Kurotaki, H., Tanaka, S., Hokamoto, K., Ochiai, R., Konishi, S., et al. (2021). Bonding strength evaluation of explosive welding joint of tungsten to ferritic steel using ultra-small testing technologies. Mater. Sci. Eng. A *826*, 141995. https://doi.org/10.1016/j.msea.2021.141995.

71. Wang, J.B., Lian, Y.Y., Feng, F., Chen, Z., Tan, Y., Yang, S., Liu, X., Qiang, J.B., Liu, T.Z., Wei, M.Y., et al. (2019). Microstructure of the tungsten and reduced activation ferritic-martensitic steel joint brazed with an Fe-based amorphous alloy. Fusion Eng. Des. *138*, 164–169. https://doi.org/10.1016/j.fusengdes.2018.11.017.

72. Computational Materials Design of Functionally Graded Structures for Enhanced Cooling of Plasma Facing Components via Additive Manufacturing | SBIR.gov https://www.sbir.gov/sbirsearch/detail/1863671.

73. Development of Reduced Activation Ferritic Martensitic (RAFM) Steels Technologies | SBIR.gov https://www.sbir.gov/node/1862539.

74. Zhong, Z., Hinoki, T., and Kohyama, A. (2010). Diffusion Bonding of Tungsten to Reduced Activation Ferritic/Martensitic Steel F82H Using a Titanium Interlayer. In Zero-Carbon Energy Kyoto 2009 Green Energy and Technology., T. Yao, ed. (Springer Japan), pp. 266–273. https://doi.org/10.1007/978-4-431-99779-5_42.

75. Wellmann, P.J. (2021). The search for new materials and the role of novel processing routes. Discov. Mater. *1*, 14. https://doi.org/10.1007/s43939-021-00014-y.

76. Razzaq, A., Allen, T.G., Liu, W., Liu, Z., and De Wolf, S. (2022). Silicon heterojunction solar cells: Techno-economic assessment and opportunities. Joule *6*, 514–542. https://doi.org/10.1016/j.joule.2022.02.009.

77. Granholm, J.M. National Blueprint for Lithium Batteries 2021-2030.

78. Bessant, J. (1996). Mastering the dynamics of innovation. Technovation *16*, 99.

79. Fitzgerald, G., Wankerl, A., and Schramm, C. (2010). Inside Real Innovation (World Scientific Publishing Company).

80. Gero, J.S., and Kannengiesser, U. (2004). The situated function–behaviour–structure framework. Des. Stud. *25*, 373–391. https://doi.org/10.1016/j.destud.2003.10.010.

81. Looney, E., Buscariolli, A., Yang, M., Raymond, G., Buonassisi, T., and Peters, I.M. (2022). Prototyping styles accelerate hardware development. Preprint at ChemRxiv, https://doi.org/10.26434/chemrxiv-2022-k4dkg https://doi.org/10.26434/chemrxiv-2022-k4dkg.

82. Cheng, K.-C., and Chen, K.-K. The impact of myopic loss aversion on continuing a troubled research and development expenditure. Afr J Bus Manage.

83. Federici, G., Biel, W., Gilbert, M.R., Kemp, R., Taylor, N., and Wenninger, R. (2017). European DEMO design strategy and consequences for materials. Nucl. Fusion *57*, 092002. https://doi.org/10.1088/1741-4326/57/9/092002.

84. Muldrew, S.I., Harrington, C., Keep, J., Waldon, C., Ashe, C., Chapman, R., Griesel, C., Pearce, A.J., Casson, F., Marsden, S.P., et al. (2024). Conceptual design workflow for the STEP Prototype Powerplant. Fusion Eng. Des. *201*, 114238. https://doi.org/10.1016/j.fusengdes.2024.114238.

85. Abolhasani, M., and Kumacheva, E. (2023). The rise of self-driving labs in chemical and materials sciences. Nat. Synth. *2*, 483–492. https://doi.org/10.1038/s44160-022-00231-0.

86. Stach, E., DeCost, B., Kusne, A.G., Hattrick-Simpers, J., Brown, K.A., Reyes, K.G., Schrier, J., Billinge, S., Buonassisi, T., Foster, I., et al. (2021). Autonomous experimentation systems for materials development: A community perspective. Matter *4*, 2702–2726. https://doi.org/10.1016/j.matt.2021.06.036.

87. Schrier, J., Norquist, A.J., Buonassisi, T., and Brgoch, J. (2023). In Pursuit of the Exceptional: Research Directions for Machine Learning in Chemical and Materials Science. J. Am. Chem. Soc. *145*, 21699–21716. https://doi.org/10.1021/jacs.3c04783.

88. Sun, S., Hartono, N.T.P., Ren, Z.D., Oviedo, F., Buscemi, A.M., Layurova, M., Chen, D.X., Ogunfunmi, T., Thapa, J., Ramasamy, S., et al. (2019). Accelerated Development of Perovskite-Inspired Materials via High-Throughput Synthesis and Machine-Learning Diagnosis. Joule *3*, 1437–1451. https://doi.org/10.1016/j.joule.2019.05.014.

89. Liu, Y., Hu, Z., Suo, Z., Hu, L., Feng, L., Gong, X., Liu, Y., and Zhang, J. (2019). High-throughput experiments facilitate materials innovation: A review. Sci. China Technol. Sci. *62*, 521–545. https://doi.org/10.1007/s11431-018-9369-9.

90. Nygren, R.E., Youchison, D.L., Wirth, B.D., and Snead, L.L. (2016). A new vision of plasma facing components. Fusion Eng. Des. *109–111*, 192–200. https://doi.org/10.1016/j.fusengdes.2016.03.031.

91. Seltzman, A.H., and Wukitch, S.J. (2020). Surface roughness and finishing techniques in selective laser melted GRCop-84 copper for an additive manufactured lower hybrid current drive launcher. Fusion Eng. Des. *160*, 111801. https://doi.org/10.1016/j.fusengdes.2020.111801.

92. Siemenn, A.E., Ren, Z., Li, Q., and Buonassisi, T. (2023). Fast Bayesian optimization of Needle-in-a-Haystack problems using zooming memory-based initialization (ZoMBI). Npj Comput. Mater. *9*, 1–13. https://doi.org/10.1038/s41524-023-01048-x.

93. Quadling, A., Lee, W.E., and Astbury, J. (2022). Materials challenges for successful roll-out of commercial fusion reactors. J. Phys. Energy *4*, 030401. https://doi.org/10.1088/2515-7655/ac73b2.

94. Zinkle, S.J., and Ghoniem, N.M. (2011). Prospects for accelerated development of high performance structural materials. J. Nucl. Mater. *417*, 2–8. https://doi.org/10.1016/j.jnucmat.2011.05.021.

95. Pavlina, E.J., and Van Tyne, C.J. (2008). Correlation of Yield Strength and Tensile Strength with Hardness for Steels. J. Mater. Eng. Perform. *17*, 888–893. https://doi.org/10.1007/s11665-008-9225-5.

96. Dennett, C.A., Buller, D.L., Hattar, K., and Short, M.P. (2019). Real-time thermomechanical property monitoring during ion beam irradiation using in situ transient grating spectroscopy. Nucl. Instrum. Methods Phys. Res. Sect. B Beam Interact. Mater. At. *440*, 126–138. https://doi.org/10.1016/j.nimb.2018.10.025.

97. Bloom, E.E., Busby, J.T., Duty, C.E., Maziasz, P.J., McGreevy, T.E., Nelson, B.E., Pint, B.A., Tortorelli, P.F., and Zinkle, S.J. (2007). Critical questions in materials science and engineering for successful development of fusion power. J. Nucl. Mater. *367–370*, 1–10. https://doi.org/10.1016/j.jnucmat.2007.02.007.

98. Guo, K., Yang, Z., Yu, C.-H., and J. Buehler, M. (2021). Artificial intelligence and machine learning in design of mechanical materials. Mater. Horiz. *8*, 1153–1172. https://doi.org/10.1039/D0MH01451F.

99. Ball, P. (2019). Using artificial intelligence to accelerate materials development. MRS Bull. *44*, 335–344. https://doi.org/10.1557/mrs.2019.113.

100. Sha, W., Guo, Y., Yuan, Q., Tang, S., Zhang, X., Lu, S., Guo, X., Cao, Y.-C., and Cheng, S. (2020). Artificial Intelligence to Power the Future of Materials Science and Engineering. Adv. Intell. Syst. *2*, 1900143. https://doi.org/10.1002/aisy.201900143.

101. Gordiz, K., Muy, S., Zeier, W.G., Shao-Horn, Y., and Henry, A. (2021). Enhancement of ion diffusion by targeted phonon excitation. Cell Rep. Phys. Sci. *2*, 100431. https://doi.org/10.1016/j.xcrp.2021.100431.